Atomistic Engineering of Phonons in Functional Oxide Heterostructures

*Seung Gyo Jeong, Ambrose Seo, and Woo Seok Choi\**

S. G. Jeong, Prof. W. S. Choi
Department of Physics, Sungkyunkwan University, Suwon 16419, Korea
E-mail: choiws@skku.edu

Prof. A. Seo
Department of Physics and Astronomy, University of Kentucky, Lexington, KY 40506, USA

Engineering of phonons, *i.e.*, collective lattice vibrations in crystals, is essential for manipulating physical properties of materials such as thermal transport, electron-phonon interaction, confinement of lattice vibration, and optical polarization. Most approaches to phonon-engineering have been largely limited to the high-quality heterostructures of III–V compound semiconductors. Yet, artificial engineering of phonons in a variety of materials with functional properties, such as complex oxides, will yield unprecedented applications of coherent tunable phonons in future quantum acoustic devices. In this study, we demonstrate artificial engineering of phonons in the atomic-scale $SrRuO_3/SrTiO_3$ superlattices, wherein tunable phonon modes were observed via confocal Raman spectroscopy. In particular, the coherent superlattices led to the backfolding of acoustic phonon dispersion, resulting in zone-folded acoustic phonons in the THz frequency domain. We could further fine-tune the frequencies from 1 to 2 THz via atomic-scale precision thickness control. In addition, we observed a polar optical phonon originating from the local inversion symmetry breaking in the artificial oxide superlattices, exhibiting emergent functionality. Our approach of atomic-scale heterostructuring of complex oxides will vastly expand material systems for quantum acoustic devices, especially with the viability of functionality integration.

# 1. Introduction

Artificial engineering of quantized lattice vibrations, *i.e.*, phonons, is a fascinating subject in both fundamental research and practical applications.[1-12] It involves creation and manipulation of new phonon modes through the synthesis of artificial crystals such as nanostructures of heterostructures with dissimilar materials. By manipulating the phonon dispersion, the group velocity, electric polarization, and density of states of the phonons could be effectively controlled.[1] Modulations of low-dimensional artifical nanostructures and crystals provide unique opportunities for controlling phonon and its related applications.[2-5] In particaulr, theses approaches will pave the way for quantum acoustic applications such as quantum Bragg mirrors and cavities,[6-8] quantum acoustic memory and transducers,[9] microwave-optical converters,[10] quantum amplifiers,[11] and circuit quantum acoustodynamics.[6] In particular, the quantized ground state phonon is envisaged as a potential resonator for managing quantum information, *e.g.*, superconducting quantum bits,[9] implying that facile engineering of the low-frequency phonons is highly necessary. However, most studies on artificial phonon engineering have been exclusively focused on III–V compound semiconductors,[7, 8, 13, 14] partially owing to their excellent crystalline qualities.

Complex perovskite oxide thin films and heterostructures provide accessible controllability of functional phenomena, which are closely associated with phonon dynamics.[15-25] Modulation of lattice structures by epitaxial strain causes a substantial shift in phonon wavelength.[17, 18] Adjustable soft-mode phonons can further lead to an enhanced dielectric constant and emergence of ferroelectricity in perovskite oxides.[19, 20] Superlattices composed of perovskite oxides have also been shown to systematically reduce the thermal conductivity or enhance electron-phonon coupling for thermoelectric purposes.[21, 22] Last but not least, phonon excitation energy in complex perovskite oxides has similar values with various quasiparticles including plasmons, excitons, phasons, magnons, skyrmions, representing a possible correlation to the phonon dynamics.[25]

Both optical and acoustic phonons can promote intriguing phonon-related phenomena in perovskite oxide heterostructures, as shown in **Figure 1**a. (i) First, artificial superlattice structures can break the local inversion symmetry.[24, 26, 27] For example, the inversion symmetry is broken from the perspective of the interfacial gray octahedral layers shown in Figure 1a, which might induce a polar optical phonon.[28] Because of the absence of inversion symmetry, such phonons are intrinsically polar. (ii) Second, acoustic phonon dispersion can be controlled by the supercell periodicity of the superlattice. The supercell structure induces backfolding of the phonon dispersion from that of the bulk, leading to a reduced Brillouin zone and tunable zone-folded acoustic phonons.

Figure 1b schematically shows the phonon dispersion of a bulk material (SrTiO$_3$, STO) and a superlattice (SrRuO$_3$/SrTiO$_3$, SRO/STO superlattice). The phonon frequency of STO is given by, $\omega_{STO}$ = ~sin($kd_{STO}$ / 2), where $d_{STO}$ and $k_{STO} = \pi / d_{STO}$ are the lattice constants of the bulk STO in the real and reciprocal spaces, respectively (black curve in Figure 1b). In contrast, the value of $k$ for a superlattice with a supercell periodicity ($d_{SL}$) of, *e.g.*, 10 $d_{STO}$, decreases by 10 times, and equivalent backfolding of the Brillouin zone appears (orange curves in Figure 1b). This leads to Raman-active zone-center acoustic phonons near the Γ-point (Figure 1c). The zone-folded acoustic phonon dispersion of the superlattice with the reduced Brillouin zone, shown in Figure 1c, was estimated using Rytov approximation as[13]

$$\cos(kd_{SL}) = \cos\left(\frac{\omega d_1}{v_{s1}}\right) \cos\left(\frac{\omega d_2}{v_{s2}}\right) - \frac{1+\alpha^2}{2\alpha} \sin\left(\frac{\omega d_1}{v_{s1}}\right) \sin\left(\frac{\omega d_2}{v_{s2}}\right), \qquad (1)$$

where $d_1$ and $d_2$ are the thicknesses and $v_{s1}$ and $v_{s2}$ are the sound velocities of the two distinct layers within the superlattices. The parameter $\alpha$ is defined as $\alpha = v_{s2}\rho_2 / v_{s1}\rho_1$, where $\rho_1$ and $\rho_2$ are the densities of the two materials, respectively. If each atomically thin layer within the superlattices well-preserve its sound velocity and density, the Rytov model would be also valid for atomically thin heterostructures. As an example, we modeled the six- and four-unit cell (u.c. ~0.4 nm) layers of the SRO and STO layers within a superlattice, *i.e.*, [6|4] superlattice ($d_{SL}$ = ~$10d_{STO}$), and used the reported $v$ and $\rho$ values of the individual SRO and STO materials.[29-32] The phonon dispersion of the superlattice exhibits the measurable peak frequencies ($\omega_{SL}$) of zone-folded acoustic phonons at ~60 cm$^{-1}$ (~1.8 THz) with a laser wavelength of 632.8 nm for Raman spectroscopy. Furthermore, it shows a tunable $\omega_{SL}$ via the atomically designed supercell thickness ($d_1 + d_2$) of the superlattices.

In this study, we demonstrated the atomic-scale precision modulation of phonon behavior (both optical and acoustic) using artificial oxide superlattices. We chose the SRO/STO superlattices as a model system to realize artificial phonon engineering. STO is an incipient ferroelectric, where the polar optical phonons are closely associated with its quantum paraelectric phenomena.[33, 34] The atomically sharp interfaces and surfaces of the SRO/STO superlattices led to the facile engineering of the phonons.[35-40] The superlattices successfully realized THz excitations of zone-folded acoustic phonon modes and largely modulated the excitation frequency up to 2 THz via atomic-scale epitaxy. Furthermore, the superlattice structure exhibited a polar optical phonon mode, which was not observed in either bulk STO or SRO, indicating local symmetry breaking. Our result demonstrates a novel route for the artificial engineering of functional phonons in complex oxide heterostructures, which would be useful for designing optical transducers in the THz region, in conjunction with the piezoelectricity and deformation potential effects.[41, 42] Furthermore, the THz frequency domain has tremendous potential for the development of next-generation communication devices due to its higher data transfer rate.[43, 44]

## 2. Results and Discussion

High-quality epitaxial SRO/STO superlattices with systematically modulated supercell thicknesses were synthesized using pulsed laser epitaxy (**Figure 2**a-2c and Figure S1-S2). We synthesized the six- and *y*-u.c. of SRO and STO layers with 50 repetitions on (001) STO substrates, *i.e.*, [6|*y*] superlattices. The number of u.c. of the superlattices was controlled utilizing a customized automatic laser pulse control system.[35, 36, 45, 46, 47] X-ray diffraction (XRD) $\theta$-$2\theta$ scans showed clear superlattice peaks ($\pm n$) with Pendellösung fringes, indicating atomically well-defined supercell structures (Figure 2a and 2b). The XRD reciprocal space maps, as shown in Figures 2c and S1, exhibited fully strained superlattices. Previous scanning tunneling electron microscopy has also shown the atomically sharp interfaces of our SRO/STO superlattices.[35, 36] The coherent supercell structures, even at 50 repetitions (thicknesses of 160-280 nm), enhance the inelastic light scattering cross-section sufficient for the experimental measurement and practical utilization of consistent phonon modes.

Precise thickness control manifests zone-folding of acoustic phonon modes with fine-tuned THz frequencies. Figure 2b shows that the angular separation between the neighboring superlattice peaks ($\Delta\theta$) systematically decreases with increasing *y*, indicating an increased $d_{SL}$. The $d_{SL}$ was characterized by Bragg's law as

$$d_{SL} = \frac{\lambda}{2}(\sin\theta_n - \sin\theta_{n-1})^{-1}, \qquad (2)$$

where $\lambda$, $n$, and $\theta_n$ are the wavelength of the X-ray (0.154 nm for Cu K-$\alpha_1$), the order of superlattice peaks, and the *n*th-order superlattice peak position, respectively. The estimated $d_{SL}$ values for [6|*y*] superlattices with *y* = 2, 4, 6, and 8 are 3.16, 3.80, 4.70, and 5.48 nm, respectively, closely matching the target thickness with a deviation smaller than 0.20 nm. Figure 2d shows the acoustic phonon dispersion of SRO/STO superlattices in the mini-Brillouin zone, modeled by Rytov approximation. With increasing *y* from 2 to 8, the estimated $\omega_{SL}$ of zone-folded acoustic phonons is systematically reduced from ~74 to ~45 cm$^{-1}$ (1.3 to 2.2 THz). This suggests that the $d_{SL}$ of the atomically well-defined superlattice is an effective control parameter for modulating the $\omega_{ph}$ of zone-folded acoustic phonons.

The theoretically estimated $\omega_{SL}$ (Figure 2d) was directly observed via confocal Raman spectroscopy, as shown in Figure 2e and Figure S3. Clear Raman excitations of zone-folded acoustic phonons were found with the modulation of their $\omega_{SL}$ systematically depending on *y*. Both Stokes and anti-Stokes shifts of the zone-folded phonons were observed for [6|8] superlattice, as shown in Figure S3. The phonons were observed in the parallel polarizations ($z(xx)\bar{z}$ or $z(x'x')\bar{z}$) but not in the cross polarizations (Figure S3), which is consistent with the previous observation of longitudinal zone-folded phonons in traditional compound semiconductors.[12] Further, we could fit the doublets of the zone-folded phonons using two Lorentzian oscillators, as shown in Figure S4, to extract $\omega_{SL}$ as the average of the doublet frequencies. The experimentally observed $\omega_{SL}$ values were in good agreement with the theoretical expectation. We summarized the $\omega_{SL}$ values of the zone-folded acoustic phonons in previous studies on various heterostructures in **Table 1**.[7, 8, 15, 23, 28, 51-56] Our result clearly shows that the $\omega_{SL}$ of atomic-scale oxide heterostructure can be largely modulated within the THz frequency range.

The linear relation between $\omega_{SL}$ and reciprocal lattice spacing ($\pi/d$) further confirms the atomistic engineering of phonons in oxide superlattices, as shown in Figure 2f. The effective sound velocity, $v_s$, estimated from the linear relation between $\omega_{SL}$ and $\pi/d$ can provide another experimental evidence. In superlattice configurations, the effective $v_s$ can be calculated by the weighted arithmetic average of the constituent layers within the supercell because the wavelength of the probing light is significantly larger than the u.c. length of the supercell. We obtained the effective $v_s$ of the SRO/STO superlattices to be approximately 6.7 nm ps$^{-1}$ (6700 m s$^{-1}$) at room temperature using a linear fit, as shown by the gray solid line in Figure 2f. We could not observe any meaningful temperature dependence of the effective $v_s$ (refer to blue empty circles in Figure 2f for $\omega_{SL}$ values at 10 K), this is also consistent with the temperature independence of $v_s$ of the SRO/STO heterostructure.[40]

**Table 2** summarizes the reference $v_s$ values of SRO and STO, and the effective experimental $v_s$ values of the SRO/STO superlattices. From the density function calculations, it was predicted that $v_s$ of bulk STO and SRO are 8.0 and 6.6 nm ps$^{-1}$,[29, 30] respectively, which coincide with the values of 7.9 and 6.3 nm ps$^{-1}$, respectively, obtained experimentally using optical spectroscopy.[31, 32] We estimated the effective $v_s$ of the SRO/STO superlattices using the following equation, considering superlattice geometry,

$$v_s = \frac{1}{4}\left(\frac{6}{8} + \frac{6}{10} + \frac{6}{12} + \frac{6}{14}\right)v_{s,SRO} + \frac{1}{4}\left(\frac{2}{8} + \frac{4}{10} + \frac{6}{12} + \frac{8}{14}\right)v_{s,STO}. \qquad (3)$$

Using the values of $v_{s,SRO}$ and $v_{s,STO}$ ($v_s$ of SRO and STO, respectively) from the previous theory and experiment (Table 2), the effective $v_s$ values of 7.1 and 6.9 nm ps$^{-1}$ were deduced. These values are in excellent agreement with our experimentally obtained effective $v_s$ of 6.7 nm ps$^{-1}$, as shown in Figure 2f. As $v_s$ represents the fundamental lattice properties of a crystal, defined by $v_s = \sqrt{c/\rho}$ (where $c$ is the elastic modulus of the material), our results also imply that the lattice properties of atomically thin SRO and STO layers within the superlattice are well preserved.

Raman spectra in the higher frequency range exhibited an unexpected phonon mode for the SRO/STO superlattice (**Figure 3**). Figure 3a compares the Raman spectrum of [6|2] superlattice with those of the SRO single film and STO substrate at room temperature. Seven Raman peaks of the STO substrate (indicated by asterisks) originate from the second-order Raman scattering of STO.[48] Five Raman peaks of the SRO single film (indicated by hashes) are the known phonon modes of the SRO layer.[45] Most of the Raman peaks of the superlattice can be interpreted as optical phonon modes of the SRO and STO layers (asterisks and hashes) or zone-folded acoustic phonon modes (~66 cm$^{-1}$) as discussed earlier. However, the two Raman excitations at ~175 and ~780 cm$^{-1}$ (diamonds) cannot be explained by the conventional Raman modes of individual SRO and STO. The surface plasmon-polariton in the SRO/STO nanoribbons revealed a similar excitation at ~800 cm$^{-1}$.[49] Yet, our measurement configuration does not allow the surface plasmon-polariton at the SRO/STO interfaces because the polarization of the incident laser is orthogonal to that of the surface plasmon-polariton. Furthermore, the temperature independence of the peak frequency, as shown in Figure 3b, does not support the surface plasmon-polariton mode.

Figure 3c schematically shows the possible polar optical phonon modes emerging in the SRO/STO superlattice, which can explain the observed Raman peaks. Whereas the original cubic STO does not host any polar phonons, TO$_2$ and LO$_4$ polar optical phonons were observed when the invension symmetry of the system breaks via various means including epitaxial strain, oxygen isotope doping, or external electric field.[23, 50, 57-59] Indeed, the peak frequencies of the Raman modes of the superlattices coinside with the emergent polar optical phonons in the STO. Moreover, the Raman spectra of the superlattice at 10 K show significant enhancements of the peaks (Figure S5), further supporting the appearance of the TO$_2$ and LO$_4$ phonons. For atomically thin superlattices, even when the ideal global inversion symmetry is preserved, local inversion symmetry can be intrinsically broken at the interfacial layers.[26, 27, 47] From density functional theory calculations, it can also be inferred that the inversion symmetry at SRO/STO interfaces is intrinsically broken.[60] Thus, we believe that the two Raman peaks at ~175 and 780 cm$^{-1}$ are TO$_2$ and LO$_4$ polar phonons of the atomically thin STO layer, originating from the local symmetry breaking of the superlattice structure.

In summary, we demonstrated atomistic engineering of phonons in deliberately designed oxide heterostructures. The atomically well-defined periodicity of the oxide superlattices led to the backfolding of acoustic phonon dispersion in the presence of zone-folded phonons in THz frequencies, which provides important implications for acoustic Bragg mirrors and cavities for generating coherent THz phonons. Furthermore, we systematically controlled the excitation energies over 2 THz via atomic-scale precision thickness control. We also observed the Raman excitation of the polar optical phonon of STO at room temperature in the superlattices. The atomically designed superlattices intrinsically break the local inversion symmetry, and thus, the polar optical phonon modes of atomically thin STO layers can be stabilized and visualized. Our approach offers a facile method for the artificial engineering of phonons in functional oxides for future quantum phononics.

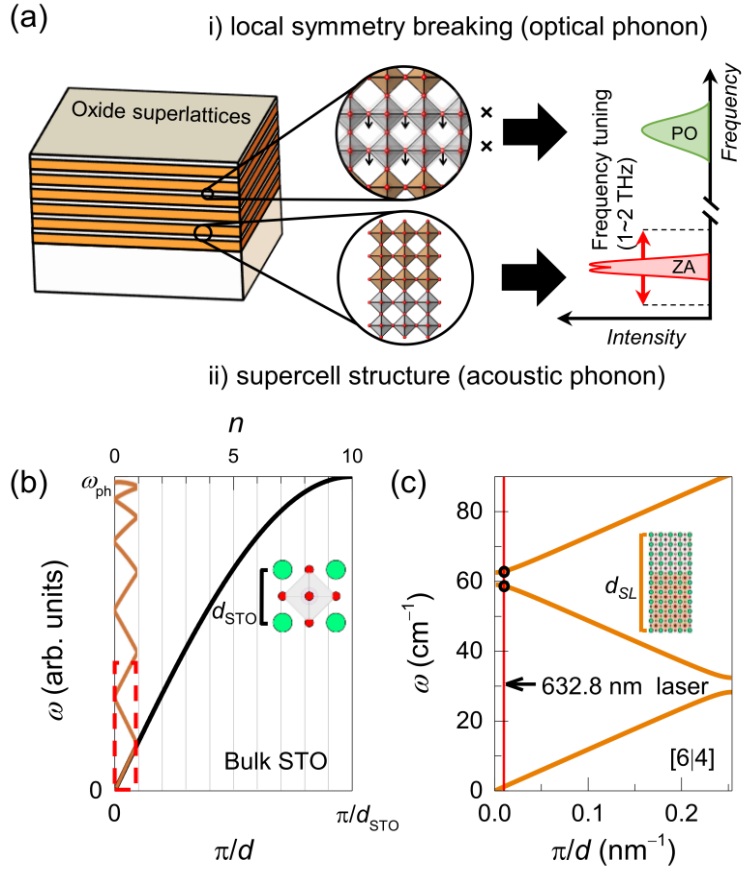

**Figure 1.** Phonon evolution in artificial oxide superlattices. a) Schematic representation of the emergent phonons in atomically designed oxide superlattices. (i) Artificial superlattice structure realizes local symmetry breaking at the interfacial layers, denoted by the crosses. As a result, new polar optical (PO) phonon modes can be stabilized and observed. (ii) The supercell structure leads to backfolding of the phonon dispersion, activating a tunable THz excitation of the zone-folded acoustic (ZA) phonon modes. b) Schematic representation of the evolution of zone-folded acoustic phonon modes in SRO/STO superlattices when $d_{SL} = 10\ d_{STO}$. The inset shows the atomic u.c. of STO with $d_{STO}$. c) Estimated acoustic phonon dispersion of [6|4] superlattice with $d_{SL}$ at low photon energy region. The vertical red line shows the $q$ value of probing laser with a wavelength of 632.8 nm.

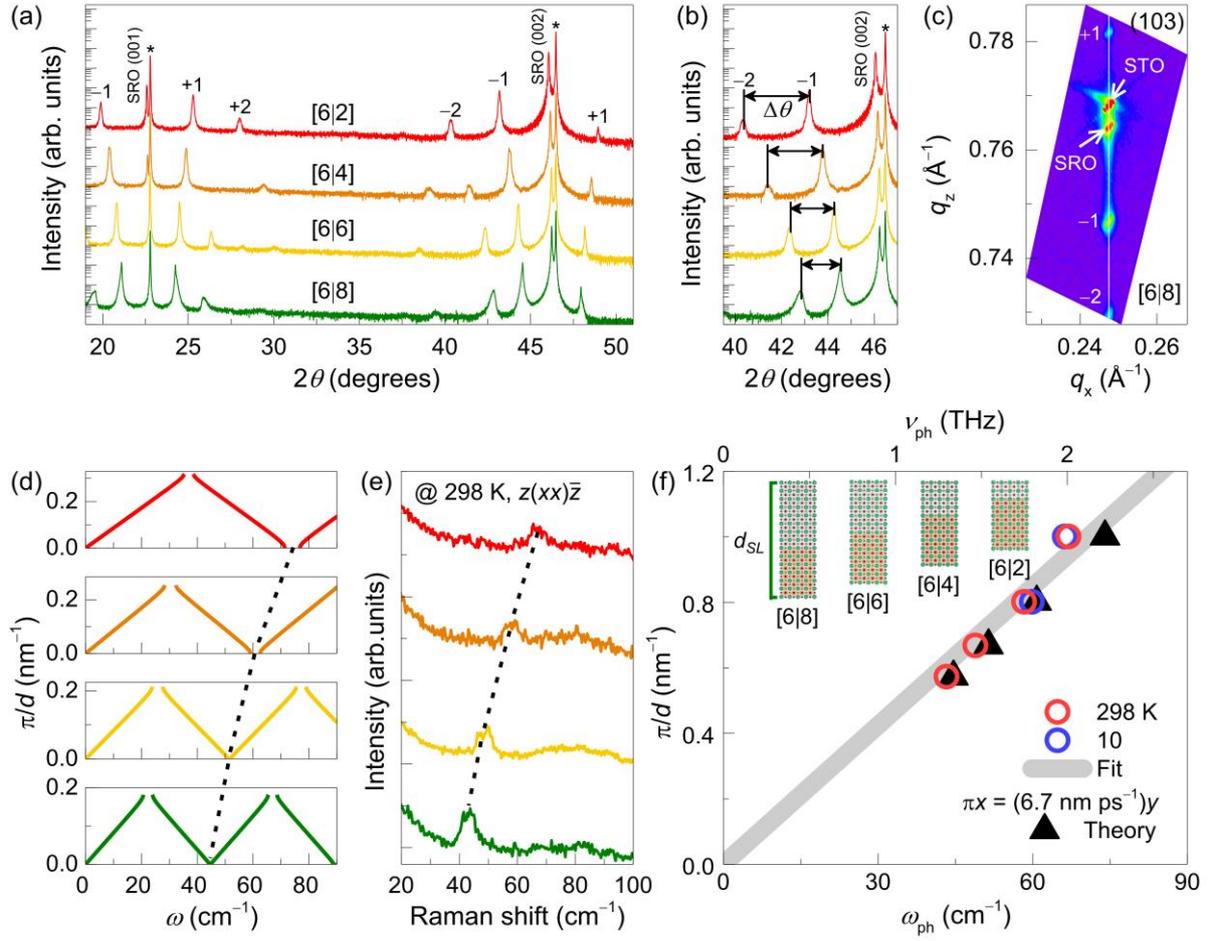

**Figure 2.** Atomic-scale phonon modulation in SRO/STO superlattices. a,b) XRD $\theta$-$2\theta$ measurements are shown for [6|y] with different y, grown on (001)-oriented STO substrates. The asterisks (*) indicate the (00l) Bragg peaks of the STO substrates. The Bragg peaks of the superlattice ($\pm n$) with $\Delta\theta$ represent well-defined periodicity. c) XRD RSM is shown for [6|8] superlattice, around the (103) Bragg reflections of the STO substrate. d) Simulated phonon dispersions and e) Raman spectra are shown for zone-folded acoustic phonon modes of superlattices with different $d_{SL}$. The spectra are measured with $z(xx)\bar{z}$ scattering configuration at room temperature. The dotted lines are shown as a the guide to the eye. f) Phonon dispersion for the low-energy region is estimated using $\omega_{ph}$ of [6|y] with different y. The solid line represents the fit of $\omega_{ph}$ at room temperature.

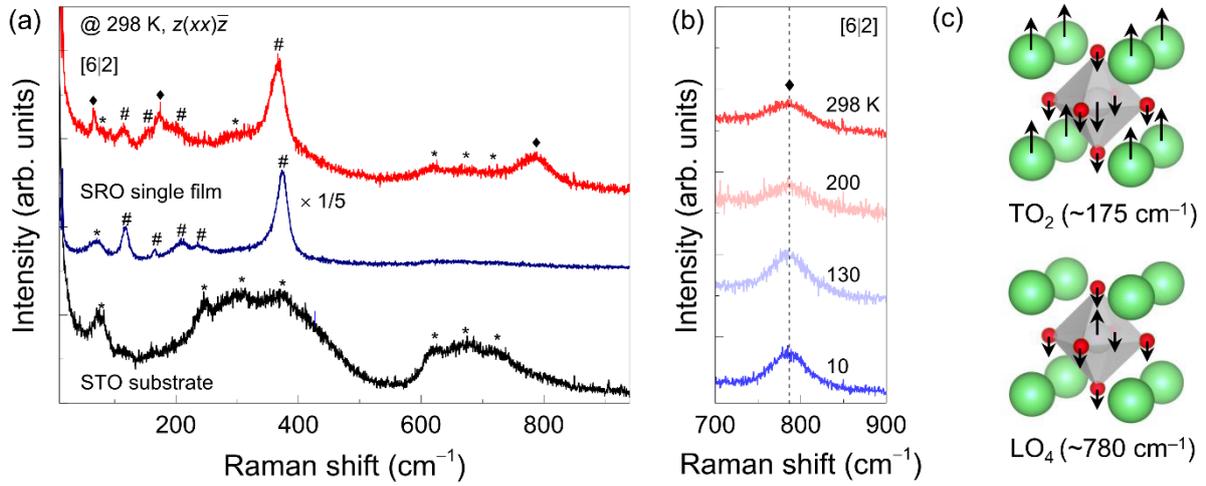

**Figure 3.** Observation of polar optical phonon mode in SRO/STO superlattices. a) Raman spectra for STO substrate, SRO single film, and superlattice. The spectra are measured with $z(xx)\bar{z}$ scattering configuration at room temperature. Raman intensity of SRO film is multiplied by 1/5. (We note that three phonon modes in the range of 600-750 cm$^{-1}$ consistently exist in the Raman spectra of the SRO single film.) The asterisk (*), hash (#), and diamond (♦) symbols indicate the phonon assignments for STO substrate, SRO layers, and superlattices, respectively. (b) Raman spectra for SRO/STO superlattice show temperature dependence of LO$_4$ mode. (c) The schematics show the polar phonon modes of TO$_2$ and LO$_4$ for TiO$_6$ octahedra within STO. The arrows indicate the eigenvector of the phonon mode along [001].

**Table 1.** Observed $\omega_{ph}$ of zone-folded acoustic phonons from previous experiments are compared to that in SRO/STO superlattice.[7, 8, 15, 23, 28, 51-56]

| Material | $\omega_{ph}$ (THz) | Repetition number | Growth method | Measurement method | Reference |
| --- | --- | --- | --- | --- | --- |
| InGaN/GaN | 0.66 – 1.23 | 14 | Metal–organic chemical vapor deposition | Time-resolved pump-probe experiments | [51] |
| GaAs/AlAs | 0.40 – 0.96 | 40 | Molecular beam epitaxy | Time-resolved pump-probe experiments | [7] |
| GaAs/AlAs | 0.50 – 0.61 | 60 or 80 | Molecular beam epitaxy | Time-resolved pump-probe experiments | [52] |
| GaAs/AlAs | 0.22 – 0.74 | 220 | Metal–organic chemical vapor deposition | Time-resolved pump-probe experiments | [53] |
| GaAs/AlAs | 0.15 – 0.90 | 11 | Molecular beam epitaxy | Raman spectroscopy | [8] |
| InSe/hBN | 0.02 – 0.15 | 1 | mechanical exfoliation | Time-resolved pump-probe experiments | [54] |
| $BaTiO_3/SrTiO_3$ | 0.36 – 0.90 | 30 or 50 | Pulse laser deposition | Raman spectroscopy | [15] |
| $BaTiO_3/SrTiO_3$ | 1.8 | 25 | Molecular-beam epitaxy | Raman spectroscopy | [28] |
| $YBa_2Cu_3O_{7-x}/La_{1/3}Ca_{2/3}MnO_3$ | 0.26 – 0.35 | 26 | Laser molecular beam epitaxy | Time-resolved pump-probe experiments | [55] |
| $YBa_2Cu_3O_7$/manganite compositions | 0.15 – 0.54 | 10 | Pulse laser deposition | Confocal Raman spectroscopy | [23] |
| $SrTiO_3/(SrRuO_3$ or $SrIrO_3)$ | 0.50 – 0.97 | 5 or 10 | Pulse laser deposition | Time-resolved pump-probe experiments | [56] |
| $SrRuO_3/SrTiO_3$ | 1.30 – 2.20 | 50 | Atomic-scale pulsed laser epitaxy | Confocal Raman spectroscopy | This work |

**Table 2.** Sound velocities of SRO and STO from previous studies are compared to that in SRO/STO superlattice.[29-32]

|  | Material | $d_{pc}$ (nm) | $u_{sound}$ (nm ps$^{-1}$) | Method | Reference |
|---|---|---|---|---|---|
| Theory | SrTiO$_3$ | 0.3940 | 8.0 | Density functional theory | [29] |
|  | SrRuO$_3$ | 0.3914 | 6.6 | Density functional theory | [30] |
| Experiment | SrTiO$_3$ | 0.3905 | 7.9 | Brillouin spectroscopy | [31] |
|  | SrRuO$_3$ | 0.3937 | 6.3 | Ultrasonic pulse-echo method | [32] |
| This work | SrRuO$_3$/SrTiO$_3$ superlattices | Systematically controlled | 6.7 | Confocal Raman spectroscopy | - |

## 4. Experimental Section

*Atomic-scale epitaxial growth and lattice characterization*: We synthesized the SRO/STO superlattices with six- and *y*-u.c. of the SRO and STO layers, *i.e.*, [6|*y*] superlattice, at 750 °C in 100 mTorr of oxygen partial pressure using pulsed laser epitaxy. To enhance the Raman cross-section of inelastic light scattering, we used the SRO/STO superlattices with 50 repetitions. We ablated stoichiometric SRO and STO targets using a KrF laser (248 nm, IPEX868, Lightmachinery) with a laser fluence of 1.5 Jcm$^{-2}$ and a repetition rate of 5 Hz. X-ray $\theta$-$2\theta$, off-axis, and reciprocal space map measurements were performed using a high-resolution PANalytical X'Pert X-ray diffractometer. X-ray rocking curve measurements show the excellent crystallinity of the superlattices even after 50 repetitions (Figure S2).

*Confocal Raman spectroscopy*: We recorded the Raman spectra of SRO/STO superlattices using a confocal micro-Raman (Horiba LabRam HR800) spectrometer with a 632.8 nm (1.96 eV) HeNe laser. Temperature-dependent measurements were performed under vacuum using an optical cryostat. We used a grating with 1800 grooves per mm and a focused beam spot with a size of ~5 μm. The power of the laser was kept below ~0.3 mW to suppress any heating effects. We deliberately controlled the *z*-directional beam position to achieve optimal focus on the superlattice samples for measuring high-quality Raman spectra using a backscattering geometry.[61] We recognize that the previous Raman spectroscopy required the resonant wavelength of the pump laser. However, confocal Raman spectroscopy let us probe the sensitive Raman intensity in atomically thin oxide layers.[19, 48] We also note that the coherent supercell structures, even up to 50 repetitions (thicknesses of 160-280 nm), increase the inelastic light scattering cross-section sufficient for the Raman intensity of the zone-folded acoustic phonon modes.

**Conflict of Interest**

The authors declare no conflict of interest.

**Supporting Information**

Supporting Information is available from the Wiley Online Library or from the author.

**Acknowledgments**

This work was supported by the Basic Science Research Program through the National Research Foundation of Korea (NRF-2021R1A2C2011340). A.S. acknowledges the support from the National Science Foundation (DMR-1454200) and the Alexander von Humboldt Foundation (Research Fellowship for Experienced Researchers).

# Supporting Information

**Atomistic Engineering of Phonons in Functional Oxide Heterostructures**

Seung Gyo Jeong, Ambrose Seo, and Woo Seok Choi*

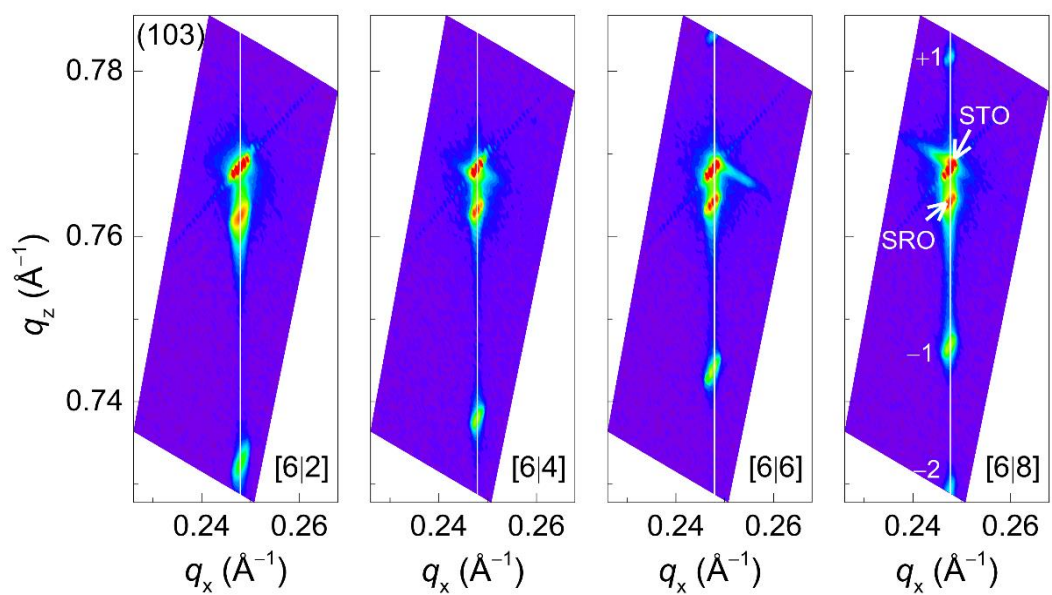

**Figure S1**. XRD RSMs of the SRO/STO superlattices. RSMs of the [6|y] superlattices with different y were measured around the STO (103) Bragg reflection. The Bragg peaks of the superlattices (±n) correspond to atomically-well defined periodicities of superlattices. The vertical lines indicate a fully strained state of superlattices.

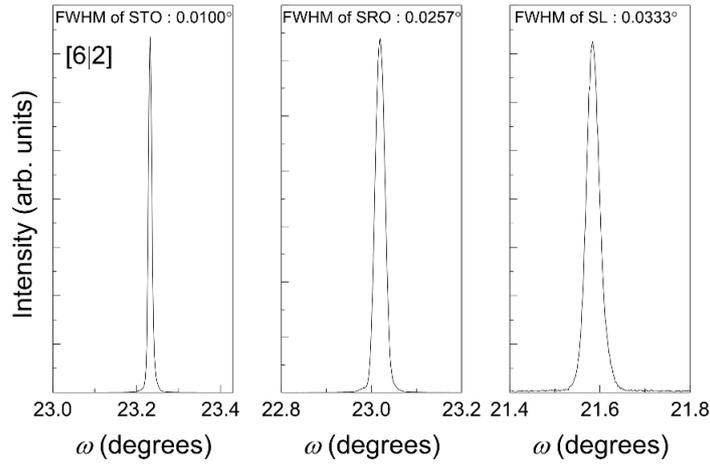

**Figure S2**. XRD rocking curve measurements of the SRO/STO superlattice. Rocking curve measurements of [6|8] superlattice were recorded around (002) Bragg reflection of each layer. The full-width-at-half-maximum (FWHM) value of SRO and superlattice peaks are comparable to that of the STO substrate. This result shows the excellent crystallinity of our oxide superlattices.

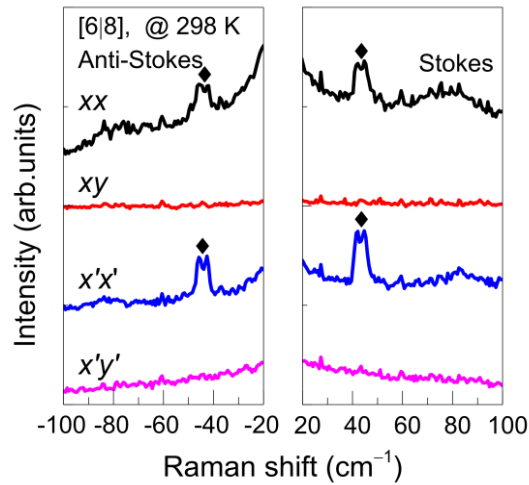

**Figure S3**. Polarization-dependent Raman spectra of the SRO/STO superlattices. Polarized Raman spectra of [6|8] superlattices were measured at room temperature. Diamonds (♦) indicate the zone-folded acoustic phonons. Both Stoke and anti-Stoke scattering assure the existence of zone-folded acoustic phonon mode in the superlattice. Polarization dependence of the phonon is consistent with previous longitudinal zone-folded phonon.[12]

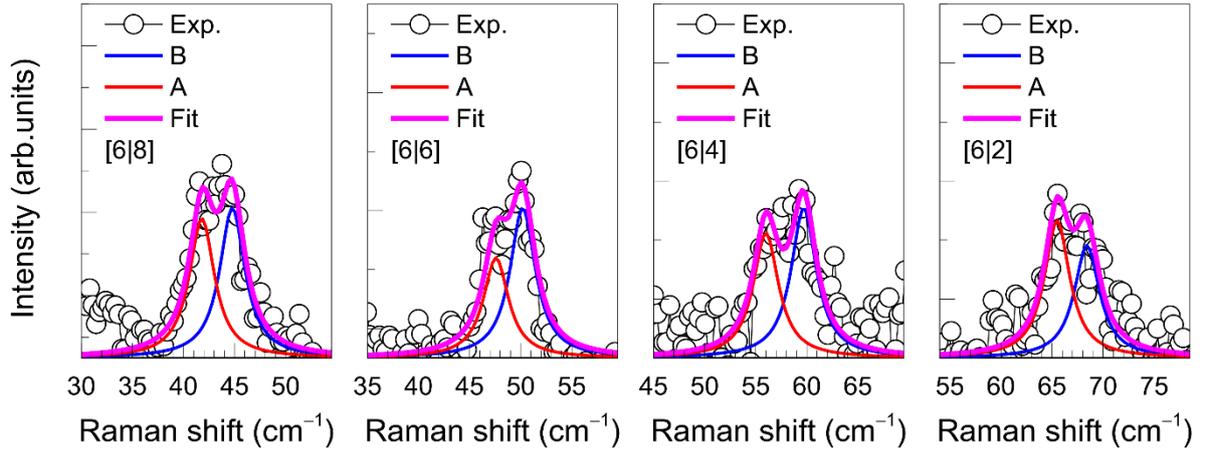

**Figure S4**. Lorentzian fitting for zone-folded acoustic phonon modes. The *y*-dependent doublet of folded acoustic phonon $\omega_{SL}$s of [6|*y*] superlattices were obtained by Lorentzian fitting using two peaks.

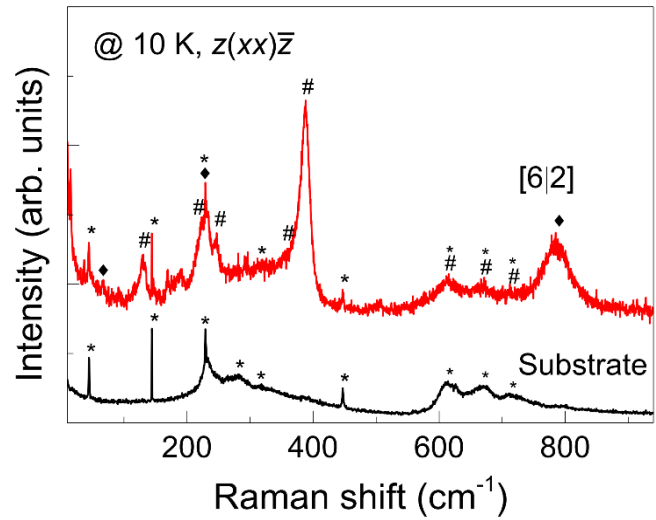

**Figure S5**. Raman spectra of the SRO/STO superlattice and STO substrate, measured at 10 K. The asterisk (*), hash (#), and diamond (♦) symbols indicate the phonon assignments for STO, SRO layers, and superlattices, respectively.